\documentstyle[aps,prb,epsf]{revtex}

\begin{document}
\twocolumn[\hsize\textwidth\columnwidth\hsize\csname@twocolumnfalse\endcsname
\author{A.E. Koshelev$^a$ and L.N. Bulaevskii$^b$}
\address{$^a$ Materials Science Division, Argonne National Laboratory, 
Argonne, Illinois 60439 \\
$^b$ Los Alamos National Laboratory, Los Alamos, NM 87545}
\title{Fluctuation broadening of plasma resonance line in the vortex
 liquid state of layered superconductors}
\date{\today}
\maketitle

\begin{abstract}  
The Josephson plasma resonance (JPR) provides a sensitive probe of vortex 
states in layered superconductors.  We demonstrate that in the case of weak 
damping in the liquid phase, broadening of the JPR line is caused mainly 
by random Josephson coupling arising from the density fluctuations of 
pancake vortices.  In this case the JPR line has the universal shape, which is 
determined only by parameters of the superconductors and temperature.  This 
mechanism gives a natural explanation for the experimentally observed 
asymmetric lineshape.  The tail at high frequencies arises due to mixing of 
the propagating plasma modes by random Josephson coupling, while the tail 
at small frequencies is caused by the localized plasma modes originating 
from a rare fluctuation suppression of the Josephson coupling in large 
areas.
\end{abstract}

\vskip.2pc] 
\narrowtext 

Observation of the Josephson plasma resonance (JPR) in the mixed state of 
layered high temperature and organic superconductors 
\cite{oph,mats,kadow,shib} allows the study of Josephson coupling of layers 
over a wide range of fields and temperatures.  The Josephson coupling in 
the mixed state is mainly determined by correlations between arrangements 
of pancake vortices in neighboring layers and therefore it can serve as a 
probe of vortex state.  The resonance absorption is determined by the 
spatial and temporal behavior of the ``local coherence parameter'' ${\cal 
C}_{n}({\bf r},t)\equiv \cos \varphi_{n,n+1}({\bf r},t)$, with 
$\varphi_{n,n+1}=\varphi_{n+1}-\varphi_{n}-(2\pi s/\Phi_{0})A_{z}$ being 
the gauge-invariant phase difference between the layers $n$ and $n+1$.  
Here ${\bf r}=(x,y)$ and $z$ are the coordinates in the $ab$ plane and 
along the $c$ axis, $s$ is the interlayer spacing.  Nonzero 
$\varphi_{n,n+1}$ arises due to the misalignment of pancake vortices in 
neighboring layers.  Time variations of $\varphi_{n,n+1}$ caused by pancake 
motion are usually much slower than the plasma oscillations.  In this case 
the plasma mode probes a snapshot of the instantaneous phase distribution.  
In the liquid phase at high fields, $B\gg B_J$, correlations between 
pancake positions in neighboring layers are almost absent.  Here 
$B_{J}=\Phi_{0}/\lambda_{J}^2$, $\lambda_J=\gamma s$ is the Josephson 
length, and $\gamma$ is the anisotropy factor.  In this state ${\cal 
C}_{n}({\bf r})$ rapidly oscillates in space so that its variations are 
much larger than average.  The resonance in such a situation occurs because 
small phase oscillations induced by the external electric field change 
slowly in space and average out these rapid variations.  We will estimate 
that the resonance is formed at the typical scale $\lambda_J^2/a$, with $a$ 
being the intervortex spacing.  This large scale averaging leads to a 
fairly sharp resonance, with the resonance frequency squared $\omega_p^2$ 
being roughly proportional to the averaged cosine factor ${\cal C}\equiv 
{\cal C}({\bf B},T)$ (see, e.g., Refs.~\onlinecite{bdmb,koshPRL96})
\begin{equation}
\omega_p^2(B,T)\approx \omega_{0}^2(T){\cal C}, \;\;\;\;
{\cal C}\equiv \langle \cos \varphi_{n,n+1}({\bf r})\rangle	
\label{omegap}
\end{equation}
where $\omega_{0}^2(T)=c^2/\epsilon_{c}\lambda_{c}^2(T)$, $\lambda_{c}$ is 
the $c$-component of the London penetration depth, $\epsilon_{c}$ is the 
dielectric constant, and $\langle \ldots \rangle $ denotes the 
thermodynamic average.  However, the averaging by the smooth oscillating 
phase is not complete.  Large scale fluctuations of ${\cal C}_{n}({\bf 
r})$, induced by pancake density fluctuations, lead to inhomogeneous 
broadening of the JPR line.  This mechanism was proposed in Ref.\ 
\onlinecite{bdmb} to explain line broadening in the vortex glass state.  In 
this paper we analyze broadening of the JPR line in the vortex liquid due 
to the random Josephson coupling.  An attractive feature of the line 
broadening in the liquid phase is that it is an intrinsic property of the 
material caused by thermal fluctuations.  The plasma resonance line has a 
very peculiar asymmetric shape with the long tail in the high field part 
\cite{oph,mats,kadow}.  This shape can be naturally explained by the 
proposed mechanism.  The high frequency/high field tail of the line probes 
mixing of the propagating plasma modes by the random Josephson coupling, 
while the low frequency/low field tail probes the localized plasma modes 
originating due to rare fluctuation suppression of the coupling.  We will 
analytically calculate the resonant absorption in both regions.

The external alternating electric field applied along the $c$ axis ${\cal 
D}_{z}={\cal D}_{z0}\exp(i\omega t)$ induces small oscillations of the 
interlayer phase difference $\varphi_{n,n+1}^{\prime }$.  In general, 
charging of layers \cite{koy} and deviations of the quasiparticle 
distribution function from equilibrium \cite{ryn} should be accounted 
for in the time-dependent equation for the phase difference.  To 
illustrate the physics of the inhomogeneous line broadening, we will consider 
a simplified equation, in which we do not account for these effects 
\cite{zam,bdmb}
\begin{equation} 
\left( \frac{\omega^2}{\omega_{0}^2}+\lambda 
_{J}^2\hat{L}\nabla^2-{\cal C}_{n}({\bf r})\right) \varphi 
_{n,n+1}^{\prime }=-\frac{ i\omega {\cal D}_{z}}{4\pi J_{0}},  
\label{DynEq} 
\end{equation} 
where
$J_{0}=c\Phi_{0}/8\pi^2\lambda_{c}^2s$ is the Josephson 
current.
The second term in the brackets accounts for the inductive interaction 
between the junctions.  The operator $ \hat{L}$ acts on the layer index $n$ 
and is defined as $\hat{L} A_{n}=\sum_{m}L_{nm}A_{m}$ with 
$L_{nm}=\int_{0}^{2\pi }\frac{dq}{2\pi } L(q)\cos (n-m)q$, and $L(q)=\left[ 
2(1-\cos q)+s^2/\lambda_{ab}^2\right]^{-1}$, with $\lambda_{ab}$ being the 
$ab$-components of the London penetration depth.  We neglect in 
Eq.~(\ref{DynEq}) time variations of ${\cal C}_{n}({\bf r},t)$ assuming 
them to be small during the time $1/\omega$.  If we neglect the charging 
effects \cite{koy}, then the oscillating internal electric field is 
connected with the solution of Eq.~(\ref{DynEq}) by the Josephson relation 
$E_{z}\approx (i\omega \Phi _{0}/ 2\pi cs)\varphi_{n,n+1}^{\prime }$.  The 
resonant absorption is given by the imaginary part of the inverse 
dielectric constant $1/\varepsilon_{c}(\omega )=E_{z}/{\cal D}_{z}$.  We 
split ${\cal C}_{n}({\bf r})$ into the average and fluctuating parts, 
${\cal C}_{n}({\bf r})={\cal C}+u_{n}({\bf r})$.  The correlation function 
of $u_{n}({\bf r})$ is given by
\[
\left\langle u_{n}({\bf r})u_{n^{\prime }}({\bf r}^{\prime
})\right\rangle=\left\langle\cos
\varphi_{n,n+1}({\bf r})\cos\varphi_{n',n'+1}({\bf r}')\right\rangle-{\cal C}^2.
\]
This correlation function depends weakly on $E_J$. Taking the limit $E_J\rightarrow 0$, 
we obtain 
\[
\left\langle u_{n}({\bf r})u_{n^{\prime }}({\bf r}^{\prime
})\right\rangle
\approx \frac{1}{2}S({\bf r}-{\bf r}^{\prime })\delta_{nn^{\prime }},
\]
where 
$S({\bf r})=\langle \cos [\varphi_{n,n+1}({\bf r})-
\varphi_{n,n+1}(0)]\rangle  $
is the static phase correlation function at $E_{J}=0$. Static configurations
$\varphi_{n,n+1}({\bf r})$ are mainly determined by the thermal
fluctuations of pancake vortices, and $S({\bf r})$ drops at distances of the order of
the intervortex spacing $a$. 

An important observation is that in spite of the rapid variations of ${\cal 
C }_{n}({\bf r})$ in Eq.~(\ref{DynEq}), the solution $\varphi^{\prime 
}\equiv \varphi_{n,n+1}^{\prime }(r)$ varies smoothly in space.  At $\omega 
\approx \omega_p$ the typical length scale $L_{\varphi }$ of phase 
variations can be estimated by balancing the typical kinetic energy of 
supercurrents, $E_{0}(\varphi^{\prime })^2/L_{\varphi }^2$, with the 
typical fluctuation of the random Josephson energy, $E_{J}(\varphi ^{\prime 
})^2a/L_{\varphi }$.  Here $E_{0}=s\Phi_{0}^2/16\pi^{3}\lambda_{ab}^2$ is 
the in-plane phase stiffness and $E_{J}=E_{0}/\lambda_{J}^{2}$ is the 
Josephson energy per unit area.  This gives $L_{\varphi 
}=\lambda_{J}^2/a\gg a$.  Smoothly varying $\varphi^{\prime }$ effectively 
averages the rapid variation of ${\cal C}_{n}({\bf r})$ and the plasma 
frequency is simply determined by $ {\cal C}({\bf B},T)$.  Fluctuations of 
${\cal C}_{n}({\bf r})$ smoothened over the area $L_{\varphi }^2$, ${\cal 
C}_{L_{\varphi }}=L_{\varphi }^{-2}\int_{r<L_{\varphi }}d{\bf r}{\cal 
C}_{n}({\bf r})$, produce inhomogeneous broadening of the JPR line.  
Calculating the mean squared fluctuation of $ {\cal C}_{L_{\varphi }}$, 
$\langle ({\cal C}_{L_{\phi }}-{\cal C} )^2\rangle \approx a^2/L_{\phi 
}^2=a^{4}/\lambda_{J}^{4}$, we obtain the estimate for the inhomogeneous 
line width
\begin{equation}
\omega_b^2\approx \omega_{0}^2(T)B_{J}/B.
\label{linew}
\end{equation}

Inhomogeneous line broadening is determined by the amplitude of 
fluctuations of $u_{n}({\bf r})$, $S_{0}=\int d{\bf r}S({\bf r})\approx 
a^2$, i.e., $\omega_{b}^{2}\propto S_{0}$.  On the other hand, according to 
the high temperature expansion \cite{koshPRL96,kbmPRL98}, the average 
cosine is also determined by $S_0$:
\begin{equation}
{\cal C}({\bf B},T)\approx (E_{J}/2T)S_{0},
\label{cfun}
\end{equation}
This relation allows us to connect the linewidth with the resonance 
frequency.  In Ref.\ \onlinecite{kbmPRL98} $S_{0}$ and ${\cal C}({\bf 
B},T)$ were calculated taking into account both the vortex and regular 
phase fluctuations.  At low temperatures, $T \ll 2\pi E_{0}$, the result 
can be written as
\begin{equation}
{\cal C}({\bf B},T)\approx \frac{f_{s} E_0B_{J}}{2TB}\left( 1+\frac{T}{2\pi E_{0}} 
\ln\left(\frac{B}{B_{J}}\right) \right),
\label{cosht}
\end{equation}
where $f_{s}(T)$ is the universal function of $T/E_{0}$.  The logarithmic 
correction appears due to the regular phase fluctuations.  Comparing 
Eqs.~(\ref{omegap}) and (\ref{cosht}) with Eq.~(\ref{linew}), we obtain an 
estimate for the relative line width in the liquid state due to the 
inhomogeneous broadening, $\omega_b/\omega_p\approx T/E_{0}\ll 1$.

Consider now the problem quantatively.  Eq.\ (\ref{DynEq}) is similar to 
the Schr\"{o}dinger equation for a particle in a random potential and the 
problem is mapped onto the problem of the density of states in such a 
system.  Then one can use the same techniques.  The problem of density of 
states does not have an analytical solution over the whole energy range.  
Only asymptotics for positive and negative energies can be calculated.

At frequencies well above the resonance frequency one can treat the random 
coupling $u_{n}({\bf r})$ as a perturbation.  A perturbative analyses of 
Eq.(\ref{DynEq}) can be conveniently performed using the Green function 
formalism.  We define the Green function ${\cal G} _{nn^{\prime }}\left( 
{\bf r},{\bf r}^{\prime };E\right) $ as a solution of the equation,
\[
\left[ E+\lambda_{J}^2\hat{L}\nabla^2-u_{n}({\bf r})\right] {\cal G}
_{nn^{\prime }}\left( {\bf r},{\bf r}^{\prime };E\right) =\delta
_{nn^{\prime }}\delta \left( {\bf r}-{\bf r}^{\prime }\right) .
\]
Here we introduced the dimensionless ``energy'' $E 
=\omega^2/\omega_0^2-{\cal C}$.  Knowledge of ${\cal G}_{nn^{\prime 
}}\left( {\bf r},{\bf r}^{\prime};E\right) $ allows one to calculate the phase 
responses to arbitrary external perturbations.  In particular, the response 
to the homogeneous external field is given by the averaged Fourier 
transform of ${\cal G} _{nn^{\prime }}\left( {\bf r},{\bf r}^{\prime 
};E\right) $,
\[
{\cal G}\left( {\bf k},q;E\right) =\sum_{n}\int d{\bf r}\left\langle
{\cal G}
_{n0}\left( {\bf r}, 0;E\right) \right\rangle \exp \left( -i{\bf kr}
-iqn\right) ,
\]
at ${\bf k},q=0$.  The dielectric constant $\varepsilon_{c}(\omega )$ 
is connected with ${\cal G}\left( {\bf k},q;E\right)$ as $ 
\varepsilon_{c}/\varepsilon_{c}(\omega )=(\omega^2/\omega_0^2) {\cal 
G}\left( 0;\omega^2/\omega_0^2-{\cal C}\right).  $ The JPR line, 
given by the imaginary part of $1/\varepsilon_{c}(\omega )$,  can 
be connected with the spectral density $A_{0}(E)$, $A_{0}(E)=
{\rm Im}\left[ \left\langle {\cal G}\left( 0;E-i\delta \right) 
\right\rangle \right]/\pi $ 
\begin{equation}
p(\omega )\equiv{\rm Im}\left[
\frac{\varepsilon_{c}}{\varepsilon_{c}(\omega )}\right] =
\frac{\pi \omega^2}{\omega_0^2}A_{0}\left( \frac{\omega^2}{
\omega_0^2}-{\cal C}\right)   .
\label{ImEps}
\end{equation}
The perturbative expansion of ${\cal G}\left( {\bf k},q;E\right) $ with
respect to $u_{n}({\bf r})$ can be performed using the diagrammatic
technique and ${\cal G}\left( {\bf k} ,q;E\right)$ is represented as
\begin{equation}
{\cal G}\left( {\bf k},q;E\right) =\frac{1}{E-L(q)\lambda
_{J}^2k^2-\Sigma \left( {\bf k},q;E\right) },  
\label{GreenF}
\end{equation}
where $\Sigma \left( {\bf k},q;E\right) $ is the self energy function.  It, 
in turn, can be represented as the perturbation series with respect to $u_n({\bf 
r})$.  In the lowest order with respect to $u_{n}({\bf r})$, which is 
equivalent to the Born approximation for scattering, $\Sigma \left( {\bf 
k},q;E\right)$, is given by
\begin{equation}
\Sigma \left( {\bf k},q;E\right) =\frac{1}{2}\int \frac{d{\bf 
k}^{\prime}dq^{\prime }}{(2\pi)^{3}}
{\cal G}\left({\bf k}^{\prime },q^{\prime };E\right) S(
{\bf k}-{\bf k}^{\prime })
\label{SigBorn}
\end{equation}
The imaginary part of $\Sigma \left( {\bf k},q;E\right)$, which we denote 
by $\Sigma_{2}$, determines the line broadening, while the real part, 
$\Sigma_{1}$, determines the shift of the resonance frequency due to 
inhomogeneities.  The high frequency asymptotics of $\Sigma \left( {\bf 
k},q;E\right)$ can be obtained by replacing ${\cal G} \left( {\bf 
k},q;E\right)$ with its bare value ${\cal G}_{0}\left( {\bf k},q;E\right) 
=\left[E-L(q)\lambda_{J}^2k^2-i\delta\right]^{-1}$.  The main contribution 
to $\Sigma_{2}$ comes from the region $k\sim \sqrt{E}/\lambda_J\ll 1/a$.  In this 
region one can neglect the $k$-dependence of $S({\bf k})$ and replace it by 
$S_0$.  The same replacement is possible for almost all terms in the 
perturbation series.  The only exception is the Born integral for 
$\Sigma_{1}$ logarithmically diverging at $k\gg \sqrt{E}/\lambda_J$.  This 
divergency is cut by the $k$ dependence of $S({\bf k})$ at $k\approx 1/a$.  
Using the replacement $S({\bf k})\rightarrow S_{0}\approx (2T/E_{J}){\cal 
C}$ and performing integrations with respect to ${\bf k}$ and $q$ we obtain 
the following expressions for the imaginary and real parts of the self 
energy function at large ``energies''
\begin{equation}
\Sigma_{2}|_{E\rightarrow \infty} \equiv \Sigma_{\infty}= \frac{T{\cal 
C}}{2E_{0}} ,\ \ \Sigma_{1}|_{E\rightarrow \infty}\approx 
-\frac{\Sigma_{\infty}}{\pi}\ln \frac{\lambda_{J}^2}{a^2E }.
\label{SelfEnBorn}
\end{equation} 
$\Sigma_{\infty}\approx a^2/\lambda_J^2$ provides the typical ``energy'' 
scale for the broadening.  In $\Sigma_{1}$ we separate the logarithmic 
contribution, $-(2\Sigma_{\infty}/\pi)\ln(\lambda_{J}^2/ a^2)$, coming from 
large $k$, and the remaining part which is determined by $S_{0}$.
Further analysis of the perturbation expansion shows that 
the expansion parameter at large $E$ is 
$\Sigma_{\infty}/[E+(2\Sigma_{\infty}/\pi)\ln(\lambda_{J}^2/a^2)]$.  
Therefore the components of the self energy can be represented in the 
following scaling form
\begin{equation}
\frac{\Sigma_{2}}{\Sigma_{\infty}}= s_2(\zeta),\ \
\frac{\Sigma_{1}}{\Sigma_{\infty}}=-\frac{2}{\pi}\ln
\frac{\lambda_{J}^2}{ a^2}-s_1(\zeta)
\label{SelfEnSc}
\end{equation}
with $\zeta=E/\Sigma_{\infty}+(2/\pi)\ln(\lambda_{J}^2/a^2)$.  $s_1(\zeta)$ 
and $s_2(\zeta)$ are the dimensionless functions with the asymptotics, 
$s_2\rightarrow 1$ and $s_1\rightarrow (1/\pi)\ln(1/\zeta)$ at $\zeta 
\rightarrow \infty$.  Using this expressions we can represent the JPR 
absorption (\ref{ImEps}) in scaling form
\begin{eqnarray}
&&p(\omega )\omega_b^2/\omega^2 =f(\zeta), \; \; \;
\zeta=(\omega^2-\tilde{\omega}_p^2)/\omega_b^2\label{LineScal} \\
&&f(\zeta)=\frac{s_2(\zeta)}{[ \zeta+s_1(\zeta)]^2+
s_2^{2}(\zeta)},
\label{ScalFun}
\end{eqnarray}
where $\omega_b$ is the resonance width, $\omega_b^2=(T/2E_{0})\omega_p^2$, 
in agreement with the estimate (\ref {linew}).  We assume this width to be 
larger than the linewidth due to the quasiparticle dissipation.  
$\tilde{\omega}_p$ is the resonance frequency shifted by the random 
Josephson coupling, $\tilde{\omega}_p^2=\omega_p^2(1-(T/\pi 
E_0)\ln(B/B_{J})$).  This negative shift is due to the second order 
perturbative correction to the ground state ``energy'', similar to the well 
known result of quantum mechanical perturbation theory.  Note that the 
inhomogeneous correction is two times larger than the correction due to the 
regular phase fluctuations (see Eq.\ (\ref{cosht})) and has the opposite 
sign.  The high frequency tail of the resonant absorption, $\omega^2-\omega 
_p^2\gg \omega_b^2$, is given by $p(\omega )\approx 
\omega^2\omega_b^2/\left( \omega^2-\tilde{\omega}_p^2 \right)^2$.  The 
whole shape of the line is determined by the single dimensionless parameter 
$\mu=T/E_{0}$.  In particular, the width of the line is $\mu\omega_p$, the 
maximum absorption scales as $1/\mu$, and the absorption at 
$\omega^2-\omega_p^2\approx\omega_p^2$ scales as $\mu$.  Using the field 
dependence of the plasma frequency given by Eqs.\ (\ref{omegap}) and 
(\ref{cosht}) we can also present the scaling parameter $\zeta$ as a 
function of the magnetic field
\begin{equation}
\zeta=\left(B-B_{r}\right)/B_{b}+(1/\pi )\ln(B/B_{J}),
\label{zetaB}
\end{equation}
with $B_{r}= f_{s}\omega _{0}^{2}E_{J}\Phi _{0}/(2T\omega ^{2})$ and 
$B_{b}\approx (f_{s}\omega _{0}^{2}/4\omega ^{2})B_{J}$$\approx 
(T/2E_0)B_{r}$.  This allows us to obtain from Eqs.\ (\ref{LineScal}) and 
(\ref{ScalFun}) the field dependence of the resonant absorption, which is 
usually probed in the JPR experiments.

\begin{figure}
\epsfxsize=3.2in \epsffile{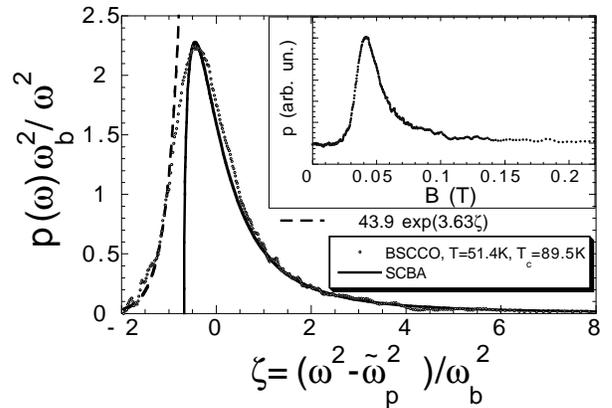} 
\caption{Plot of the reduced JPR 
lineshape $p(\omega)\omega_b^2/\omega^2$ vs reduced frequency $\zeta$ 
obtained within SCBA (Eqs.\ (\protect \ref{LineScal}) and (\protect 
\ref{SCBA})).  For comparison we also show the experimental JPR line for 
BSCCO plotted as the function of $\zeta$ from Eq.\ (\ref{zetaB}) (courtesy 
of M.~Gaifullin and Y.~Matsuda, inset shows raw data).  The parameters 
$B_{r}$ and $B_{b}$ in Eq.\ (\ref{zetaB}) are chosen to make the 
experimental and theoretical lines match at large $\zeta$ giving 
$B_{r}=0.053$T and $B_{b}=0.016$T. The left hand side of the line is fitted 
to exponent for comparison with Eq.\ (\protect \ref{exptail})}
\label{Fig-SCBorn}
\end{figure}

The problem is now reduced to a calculation of the dimensionless functions 
$s_1$ and $s_2$.  These functions can be calculated approximately if we 
keep only the first term in the expansion of $\Sigma$ given by Eq.\ 
(\ref{SigBorn}) (self consistent Born approximation (SCBA)).  This leads to 
the following equations
\begin{eqnarray}													
s_2&=&\frac{1}{2}+\frac{1}{\pi}\arctan\frac{\zeta	+s_1}{s_2} 
\nonumber \\	
s_1&=&-\frac{1}{2\pi }\ln \left[ (\zeta +s_1)^2+s_2^2\right]	,	
\label{SCBA}														
\end{eqnarray}														
which we solve numerically.  Fig.\ \ref{Fig-SCBorn} shows the dependence of 
the reduced JPR absorption $p(\omega)\omega_b^2/\omega^2$ 
on the reduced frequency $\zeta$.  Note that the SCBA actually breaks down 
at $\zeta \lesssim 1$, i.e., it describes quantitatively only the right 
hand side of the line.  For comparison we took the experimental JPR line 
for optimally doped Bi$_2$Sr$_2$CaCu$_2$O$_8$ (BSCCO)  crystal 
with $T_{c}=89.5$K \cite{matsLT21} at $\omega/2\pi=45$GHz and $T=51.4$K. We 
replotted this line as the function of the scaling parameter $\zeta$ from 
Eq.\ (\ref{zetaB}), where the parameters $B_{r}$ and $B_{b}$ are chosen to 
make the experimental and theoretical lines match at large $\zeta$ giving 
$B_{r}=0.053$T and $B_{b}=0.016$T. The obtained ratio $B_{b}/B_{r}\approx 
0.3$ is somewhat larger then the theoretical estimate $T/2E_{0}\approx 0.17$, 
which we obtain taking $\lambda_{ab}=200{\rm nm}/\sqrt{1-(T/T_{c})^{2}}$.

The SCBA approach gives the JPR line terminating at some finite frequency.  In 
reality a long absorption tail exists on the low frequency side of the line 
due to the rare fluctuation configurations corresponding to suppression of 
${\cal C}_n({\bf r})$ over large areas.  This is very similar to the 
localization tail below the bottom of the conduction band in disordered 
semiconductors (Lifshitz tail, see, e.g., Ref.~\onlinecite{LifTail}).  An 
important simplification is that due to the large scale of spatial changes 
of $\varphi_{n,n+1}^{\prime }(r)$, the random ``potential'' $u_{n}({\bf r})$ 
can be treated as a short ranged Gaussian random variable with the 
probability distribution
\[
P[u_{n}({\bf r})]\propto \exp \left[ -{\cal H}\{u_{n}\}\right] ,\ \  
{\cal H}\{u_{n}\}=\sum_{n}\int d{\bf r}\frac{\left[ u_{n}({\bf
r})\right]^2}{S_{0}}.
\]
Following a standard line of reasoning \cite {LifTail}, we estimate the 
spectral density $A_{0}(E)$ with exponential accuracy as $A_{0}(E)\propto 
\exp(-\Phi(E))$ with
\begin{equation}
\Phi (E)=\min_{u}{\cal H}\{u_{n}({\bf r})\}|_{{\cal E}_{0}\{u\}=E}, 
\label{PhiE}
\end{equation}
where ${\cal E}_{0}\{u\}$ is the ground state energy for a given potential 
fluctuation $u_{n}({\bf r})$,
\begin{equation}
{\cal E}_{0}\{u\}=\min_{\Psi }\left[ H_{0}\{\Psi\}+\sum_{n}\int d{\bf 
r}u_{n}({\bf r} )\Psi_{n}^2({\bf r})\right],
\end{equation}
with 
$H_{0}\{\Psi\}=\lambda_{J}^2\sum_{n,m}\int d{\bf r}L_{nm}\nabla 
\Psi_{n}\nabla \Psi_{m}$ and normalization $\sum_{n}\int d{\bf 
r}\Psi_{n}^2=1$.  A conditional minimization in Eq.\ (\ref{PhiE}) can be 
performed using the Lagrange technique
\begin{eqnarray}
&&\Phi (E)=-\beta E+ \label{phie-Lagr} \\
&&\min_{\Psi,u}\left[ \beta H_{0}\{\Psi\}+{\cal H}\{u_{n}({\bf r})\}
+\beta \sum_{n}\int d{\bf r}u_{n}({\bf r})\Psi_{n}^2({\bf r}) \right], \nonumber
\end{eqnarray}
where $\beta $ is the Lagrange factor, which has to be found from the 
relation ${\cal E}_{0}\{u\}=E$.  Optimization with respect to 
$u_{n}({\bf r})$ gives $ u_{n}({\bf r})=-S_{0}\beta \Psi_{n}^2({\bf 
r})/2$. Substituting this expression into Eq.\ (\ref{phie-Lagr}) and 
variating it with respect to $\Psi_{n}$, we obtain the nonlinear 
eigenvalue problem
\begin{equation}
-\lambda_{J}^2\hat{L}\nabla^2\Psi_{n}({\bf r})-(\beta 
S_{0}/2)\Psi_{n}^{3}({\bf r})=E\Psi_{n}({\bf r}),
\end{equation}
which determines the optimum $\Psi_{n}$ and $\beta$.  To simplify this 
equation we (i) apply the operator 
$\hat{L}^{-1}=s^2/\lambda_{ab}^2+\nabla^2_{n}$ on both sides and neglect 
small term $s^2/\lambda_{ab}^2$ in the resulting equation (here 
$\nabla^2_{n}$ is defined as $\nabla^2_{n}\alpha_{n}\equiv 
\alpha_{n+1}+\alpha_{n-1}-2\alpha_{n}$) and (ii) introduce dimensionless 
variables $\psi_{n}({\bf r})=\Psi_{n}({\bf r})/\Psi_{0}(0)$, $\tilde{{\bf 
r}}={\bf r}\sqrt{\beta S_{0}\Psi_{0}^2(0)/2}/\lambda_{J}$, and 
$\epsilon_{o}= 2E/(\beta S_{0}\Psi_{0}^2( 0)) $.  These transformations 
lead to the dimensionless nonlinear eigenvalue problem
\begin{equation}
-\tilde{\nabla}^2\psi_{n}+\nabla_{n}^2\psi_{n}^{3}=
-\epsilon_{o}\nabla_{n}^2\psi_{n}({\bf \tilde{r}}),  \label{DimEq}
\end{equation}
which has to be solved with the condition $\psi_{0}(0)=1$ and 
with the dipolelike asymptotics at large distances
\[
\psi_{n}({\bf \tilde{r}})=-a\nabla_{n}^2
(\left| \epsilon_{o}\right| \tilde{r}^2+n^2)^{-1/2},\ \  
\tilde{r},n\gg 1.
\]
The spectral density $A_{0}(E)$ can be expressed through $\epsilon_{o}$ and $
\psi_{n}({\bf \tilde{r}})$ as 
\begin{equation}
A_{0}(E)\propto \exp \left( -\frac{\psi_{4}\lambda_{J}^2E}{\left|
\epsilon_{o}\right| S_{0}}\right), \ \ 
\psi_{4}=\sum_{n}\int d{\bf \tilde{r}}\psi_{n}^{4}.  \label{Atail}
\end{equation}
A numerical solution of Eq.\ (\ref{DimEq}) gives $ \epsilon 
_{o}=-0.164$, $\psi_{4}=2.418$, and $a=0.224$.  From Eq.\ (\ref 
{Atail}) we now obtain $A_{0}(E)\propto \exp \left( 
-14.72\lambda_{J}^2E/S_{0} \right)$, which corresponds to 
the exponential tail of the JPR absorption
\begin{equation}
p(\omega ) \propto \exp(3.18 \zeta)= \exp
\left[ -\frac{7.36E_{0}\left( \tilde{\omega}_p^2-\omega^2\right)} 
{T\omega_p^2}\right].
\label{exptail}
\end{equation}
An exponential fit of the left hand side of the experimental line in Fig.\ 
\ref{Fig-SCBorn} gives $p\propto \exp(3.63 \zeta)$, consistent with above 
result.

In conclusion, we studied the universal shape of the JPR line due to 
thermal fluctuations of pancakes in the vortex liquid.  We showed that the 
line broadening is strongly asymmetric, in agreement with experimental 
observations.  We found that the relative linewidth is given by the 
parameter $T/E_0$, which also determines the strength of thermal 
fluctuations at $B=0$ and is directly related to the temperature dependent 
$\lambda_{ab}$.

We would like to thank M.\ Gaifullin and Y.\ Matsuda for providing data 
used in Fig.\ \ref{Fig-SCBorn}, I.\ Aranson for help in numerical 
calculations, and L.\ Paulius for help in improving the presentation.  At 
Argonne this work was supported by the NSF Office of the Science and 
Technology Center under contract No.\ DMR-91-20000, and by the U.\ S.\ DOE, 
BES-Materials Sciences, under contract No.\ W-31- 109-ENG-38.  At Los 
Alamos this work was supported by U.S.\ DOE. 

\vspace*{-10pt}

\end{document}